\documentstyle[psfig]{caosp}

\begin{document}

\htitle{Doppler--Zeeman mapping of magnetic CP Stars: solution ...}
\title{Doppler--Zeeman mapping of magnetic CP stars: solution of the inverse
problem simultaneously from the Stokes I, V, Q and U parameters}

\author{ D.V. Vasil'chenko \inst{1} \and 
         V.V. Stepanov     \inst{1} \and 
         V.L. Khokhlova    \inst{2} }

\institute{Department of Computational Mathematics and Cybernetics, Moscow
               State University, Moscow, 119899 Russia \and 
            Institute of Astronomy, Russian Academy of Sciences, 
               Pyatnitskaya ul.~48, Moscow, 109017 Russia }

\pubyear{1998} \volume{28} \firstpage{999}

\offprints {D.V.Vasilchenko, $e-mail$: demion@demion.msk.ru }

\date{}
\maketitle

\begin{abstract}
An efficient method has been developed for solving the inverse problem of
Doppler--Zeeman mapping of magnetic, chemically peculiar stars. A
regularized iteration method is used to simultaneously solve the integral
equations for the Stokes I, V, Q and U parameters. The validity of
analytical fits to the local profiles of the Stokes parameters is
substantiated. The algorithm had been tested on models and makes it possible
to obtain simultaneously, from the observed Stokes profiles, a map of the 
distribution of a chemical element and the parameters of an arbitrarily
shifted magnetic dipole.
\end{abstract}

In order to find local characteristics of the stellar surface of a rotating
magnetic chemically peculiar star from variations of the observed intensity
and polarization spectra, the so-called ill-posed inverse problem must be
solved.
The method used is based on one of regularizing algorithms, for
example, on Tikhonov's regularization (Tikhonov et al. 1990; Rice
1991) or the principle of maximum entropy (Vogt et al. 1987; Brown et al.
1991).

Initially, the method was developed and used for the mapping of chemical
elements without allowing for the magnetic field on the
stellar surface. It was repeatedly applied to stars with fairly weak magnetic
fields: $\epsilon $~UMa (Wehlau et al. 1982), $\theta $~Aur (Khokhlova et
al. 1986), 21~Per (Wehlau et al. 1991). On the other hand, codes were
developed that were used to determine the strength and configuration of the
magnetic field without a simultaneous construction of, nor allowance for a
``chemical map'' (Piskunov 1985; Donati et al. 1989).

In reality, when the magnetic field reaches some limiting strength and for a
given chemical inhomogeneity, the problems of determining the
configuration and strength of the magnetic field and of the chemical abundance
distribution must be combined, and the system of
four non-linear integral equations must
be solved simultaneously for all local Stokes I, V, Q, and U parameters.
In such a way, a chemical and magnetic map can be built:

$$
I,V,Q,U(\lambda ,\omega t)=\int_{cos\theta > 0}I,V,Q,U(M,\omega t)
u_1(\theta )u_2(\theta)dM,\eqno(1)
$$

The observed Stokes parameters at each phase are the left-hand sides of
these equations, while integrals of the local profiles of these parameters
over the visible stellar surface are the right-hand sides of the equations. The
integrands in the right-hand parts of the equations are determined by the local
values of the magnetic field and by the number of absorbing atoms of the
chemical element that form the spectral line. A mathematical model for the
formation of an absorption line must incorporate the dependence of the local
profiles on the local physical parameters that characterize the stellar
atmosphere.

In order to solve system (1), the integrands in the right-hand parts of the
equations must be explicitly written, i.e., a mathematical model for the
solution of the inverse problem must be formulated. Local profiles of the
Stokes parameters result from radiative transfer in the stellar
atmosphere in the presence of a magnetic field. These profiles can be
determined in two ways.

The first one is numerical integration of radiative transfer. Currently,
numerical integration of the transfer equation and calculation of the
specific intensity at frequencies of a spectral line as a function of
elemental abundance presents no problems, PROVIDED THE MODEL ATMOSPHERE IS
KNOWN. The pre-calculated grids of profiles for various lines as a function
of chemical composition and physical conditions are stored in the computer's
memory; in the process of iterations, data are retrieved from the tables and
interpolated.

However, the solution of the problem that involves a magnetic field is
severely complicated, because the local profiles of the Stokes V, U, and Q
parameters depend on instantaneous angles between the line of sight, the
local magnetic-field vector at each point M on the surface of a star, and
the optical axis of the analyzer, all of which vary during stellar rotation.
Huge unrealistic amounts of data should be pretabulated and interpolated
in this case. To compute directly Stokes profiles and their derivatives at
each iteration by known methods is also unrealistic.

To overcome the difficulties which arise when numerical solutions of
transfer equations are involved at each step of the iterative process, we
sought to make the most of analytical fits, which allowed us to minimize the 
number of parameters to be determined in the solution of the inverse problem
and also to find derivatives analytically. The technique of the solution is
given in more detail in the paper by Vasil'chenko et al. (1996).

The solution of the problem was made possible by the use of analytic
approximations for local line profiles (each polarized component of the Zeeman
pattern resolved in the presence of a magnetic field is treated in the same
way as the intensity profile in the absence of a magnetic field). This has
permitted us to model also multiplet lines, the He\,{\sc i} 5875 line for example.

For the local profiles we used well known analytic solutions of transfer
equations in the presence of a magnetic field (Unno, 1956, Landolfi et
al., 1982, reviewed by Jefferies et al., 1989), obtained for a Milne-Eddington
atmosphere model a with linear depth dependence of temperature. As it was
shown earlier by Hardorp et al.(1976), the numerical solution of the transfer
equation for an LTE atmosphere model gives practically the same shape of
magnetically splitted lines as the analytic solution given by Unno (1956).
To make these local profiles even more realistic, one may ``scale'' them by
finding approximating parameters from intensity profiles computed by numerical
integration of the transfer equation, for the atmosphere model assumed to be
correct in reality.

The magnetic field is determined assuming an arbitrary shifted and oriented 
dipole or a centered dipole with a coaxial quadrupole. Numerical modelling 
proved that until the quadrupole moment does not exceed half of the dipole
one, the magnetic field structure on the stellar surface may be equally well fitted by both models, as was found by Deridder et al. (1979); but for a
bigger contribution of the quadrupole, the fit by a shifted dipole starts to be inconvenient.

The non-linear system of integral equations (1) is an
ill-posed problem. To obtain a stable solution, we have to use some
sort of regularization algorithm. The unknowns in this system 
are distribution of the chemical species over the stellar surface and the
magnetic field parameters: value, orientation and shift (or quadrupole 
parameters) of the dipole. Such a complex set of unknowns requires 
most powerful inversion methods. 
We used a modification of the iterative regularized Newton method 
(Bakushinskii \& Goncharskii, 1988), with a special type of 
regularization accommodated to our problem.
This method has second order of convergence and allows us to obtain
a stable solution for a reasonable number of iterations.

Many tests of the code were performed to show its efficiency to recover 
simultaneously the magnetic field structure and the chemical map. It was
shown that for a strong magnetic field, when the Zeeman structure is
resolved in the stellar spectrum, I parameter profiles are sufficient to
recover the magnetic structure.
For weak fields over 0.2 kG, V-profiles are necessary to recover the field
configuration.

\acknowledgements
We acknowledge the support of this work by the Soros International Science
Foundation and the Government of the Russian Federation Science Foundation
(grants nos. N2L000 and N2L300) and also by the programme ``Astronomy''
(project no.3-292).

\end{document}